\documentclass[12pt]{iopart}
\usepackage{graphicx,amssymb} 
\usepackage{epsfig}
\begin{document}

\title[Neon Atoms Oscillating Inside Carbon and Boron-Nitride Nanotubes]{\bf Neon Atoms Oscillating Inside Carbon and Boron-Nitride Nanotubes: A
Fully Atomistic Molecular Dynamics Investigation}

\author{Karl M. Garcez$^1$, David L. Azevedo$^1$, Douglas S. Galv\~ao$^2$}
\address{$^1$Departamento de F\'{\i}sica, Universidade Federal do  Maranh\~{a}o, 65080-040 S\~ao
Luis, Maranh\~ao, Brazil \\$^2$ Instituto de F\'{\i}sica Gleb
Wataghin, Universidade Estadual de Campinas, CP 6165, 13083-970 Campinas, SP, Brazil.\\
\eads{david@ufma.br, galvao@ifi.unicamp.br}}

\begin{abstract}

In the present work we discuss based on extensive fully atomistic
molecular dynamics simulations the dynamics of Neon atoms
oscillating inside (5,5)single-walled carbon nanotubes (CNTs) and
boron-nitride (BNNT) ones. Our results show  that sustained high
frequency oscillatory regimes are possible for a large range of
temperatures. Our results also show that the general features of
the oscillations are quite similar to CNT and BNNT, in contrast
with some speculations in previous work literature about the
importance of broken symmetry and chirality exhibited by BNNTs.


\end{abstract}


\date{\today}

\maketitle

\section{Introduction}

Since the pioneering work of Cumings and Zetttl \cite{zettl}
carbon nanotube (CNT) based oscillators have became object of many
studies. Cumings and Zettl demonstrated the controlled and
reversible telescopic extension of multiwalled CNTs, thus
realizing ultra low-friction nanoscale linear bearings. Zheng and
collaborators \cite{zheng1,zheng2} have proposed that this
phenomenon could be exploited to built nano-oscillators in the
frequency range of gigahertz. Legoas \textit{et al.}
\cite{legoas1,legoas2} carried out molecular dynamics (MD)
simulations and concluded that sustained oscillations are possible
when the radii difference between inner and outer oscillator
moving parts are about $\simeq$ 3.4 {\AA}. Frequencies as high as
40 GHz are possible and even clear chaotic signatures were
observed \cite{coluci}. Rivera \textit{et al.} \cite{rivera1}
studied the behavior of oscillator damping via MD and they showed
that the damping depends on the inverse of the CNT length, which
is in good agreement with the prediction of analytical models
\cite{zheng1,zheng2}. They also showed \cite{rivera2} that the
frequencies in the gigahertz decrease as the length of the tubes
increases. Zhao \textit{et al.} \cite{zhao} studied the oscillator
energy dissipation energy mechanism, also using MD simulations,
and they demonstrated the importance of the overlap length of
double CNTs and the mechanical deformation of the outmost CNT to
this mechanism. In spite of the large amount of investigation of
CNT based oscillators some aspects remain not clearly understood,
especially with relation to friction mechanisms
\cite{kolmogorov,tangney,kis}. In this sense it is important to
study other related systems such as oscillators made of different
materials and also atoms and molecules oscillating inside the
tubes. Only very few studies have been reported along these lines,
such as oscillators composed of boron nitride nanotubes (BNNTs)
\cite{lee}, BNNTs in association with CNT ones \cite{kang},
$C_{60}$ molecules \cite{wang, liu} and rare-gas atoms
\cite{lin,zeng} inside CNTs. The choice for using rare-gas atoms
is associated with their inert nature which prevents chemical
reactions with the tubes. It has been demonstrated \cite{lin,zeng}
that sustained oscillations are possible for a large range of
energy variations and also, in some circunstantes, the tube
chirality is important to determine the oscillatory patterns.
However, there are conflicting data for the behavior of the
oscillator frequency decaying with the tube diameter. Further
studies are necessary to better understand these aspects. In the
present work we report extensive fully atomistic molecular
dynamics simulations for the study of Ne atoms oscillating inside
(5,5)CNTs and (5,5)BNNTs, to our knowledge this is the first study
of rare-gas atoms oscillating inside BNNTs.

\section{Methodology}

We have carried out molecular dynamics simulations using the
universal force field \textsc{uff} ~\cite{rappe,root} as
implemented in the \textit{Cerius2} package ~\cite{acc}. This
force field includes van der Waals, bond stretch, bond angle bend,
inversion, torsion and rotation terms. It has been used with
success in the study of dynamical properties of carbon based
nanostructures \cite{legoas1,legoas2,braga1,troche1,troche2}.
Initially, we perform a molecular mechanics energy minimization to
optimize the geometry of each isolated nanotube.  Then, a
\textbf{Ne} atom is placed about 1{\AA} from the nanotube ends and
the dynamical simulations are carried out. We have used in all
cases a microcanonical ensemble (\textsc{NVE}). We have used an
integration \textit{step} of 1 fs. The MD simulations were
performed in the following temperatures (in Kelvins): 5, 10, 20,
40, 80, 160, 240, and 300. In order to thermically equilibrate the
system a running of 200 ps is initially performed. After the
equilibration a running of 50 ps is carried out and the data
trajectory (positions and velocities) saved.  From the trajectory
data we perform a fast Fourier transform(\textsc{fft}) to obtain
the oscillatory frequencies. We have considered CNT and BNNT with
(5,5) chirality type. Two different lengths were considered (25
{\AA} and 50 {\AA} in order to investigate the frequency
dependence with the length of the tubes. The choice of (5,5)
nanotubes is due to that they have the minimum diameter that
allows the \textbf{Ne} atom encapsulation and also this particular
chirality implies fewer atoms in the unit cell which reduces
substantially the computational cost of the simulations.

\section{Results and Discussions}

In table \ref{t.Necnt} we show the calculated oscillation
frequencies for a \textbf{Ne} atom inside a \textsc{CNT}, for
temperatures from 5K up to 240K. For 300K we observed that the
\textbf{Ne} atom is not encapsulated. In our simulations all the
atoms of the nanotube are free to vibrate and with the increase of
temperature the amplitude of the vibrations of the atoms at the
end of tube increase. Mainly due to the van der Walls interactions
a dynamical barrier is created which prevents the \textbf{Ne} atom
encapsulation. As we can see from the Table stable frequencies are
possible for a large range of temperatures, which could be an
interesting aspect to be exploited in possible applications, such
as clocks. For the 25 {\AA} CNT from 5 up to 240 K the frequencies
are 136.71 and then they drop to 117.18 GHz. For the longer tube
(50 {\AA}), as expected, this occurs at lower temperature (160 K),
the frequencies drops from 78.12 to 58.59 GHz. This sustained
oscillatory regime is better visualized in Figs. 1 and 2. These
results are consistent with the ones obtained by Zeng \textit{et
al.} \cite{zeng} in spite of their different protocol simulations.
They used an impulse dynamics that can be mapped into our NVE
simulations varying the initial conditions with respect to the
temperature of simulations.

\begin{table}[!h]
\caption{Calculated frequencies (\textbf{GHz}) for Ne@CNT(5,5).}
\label{t.Necnt}
\begin{center}
\begin{tabular}{cccccccc}
\hline \hline
& & & Temperature (K)& & & & \\
 \hline \textbf{CNT}&5&10&20&40&80&160&240 \\
\hline 25 \AA&136.71&136.71&136.71&136.71&136.71&136.71&117.18  \\
\hline 50 \AA&78.12&78.12&78.12&78.12&78.12&58.59&58.59 \\
\hline \hline
\end{tabular}
\end{center}
\end{table}

\begin{figure}[!h]
\begin{center}
\includegraphics[scale=0.6]{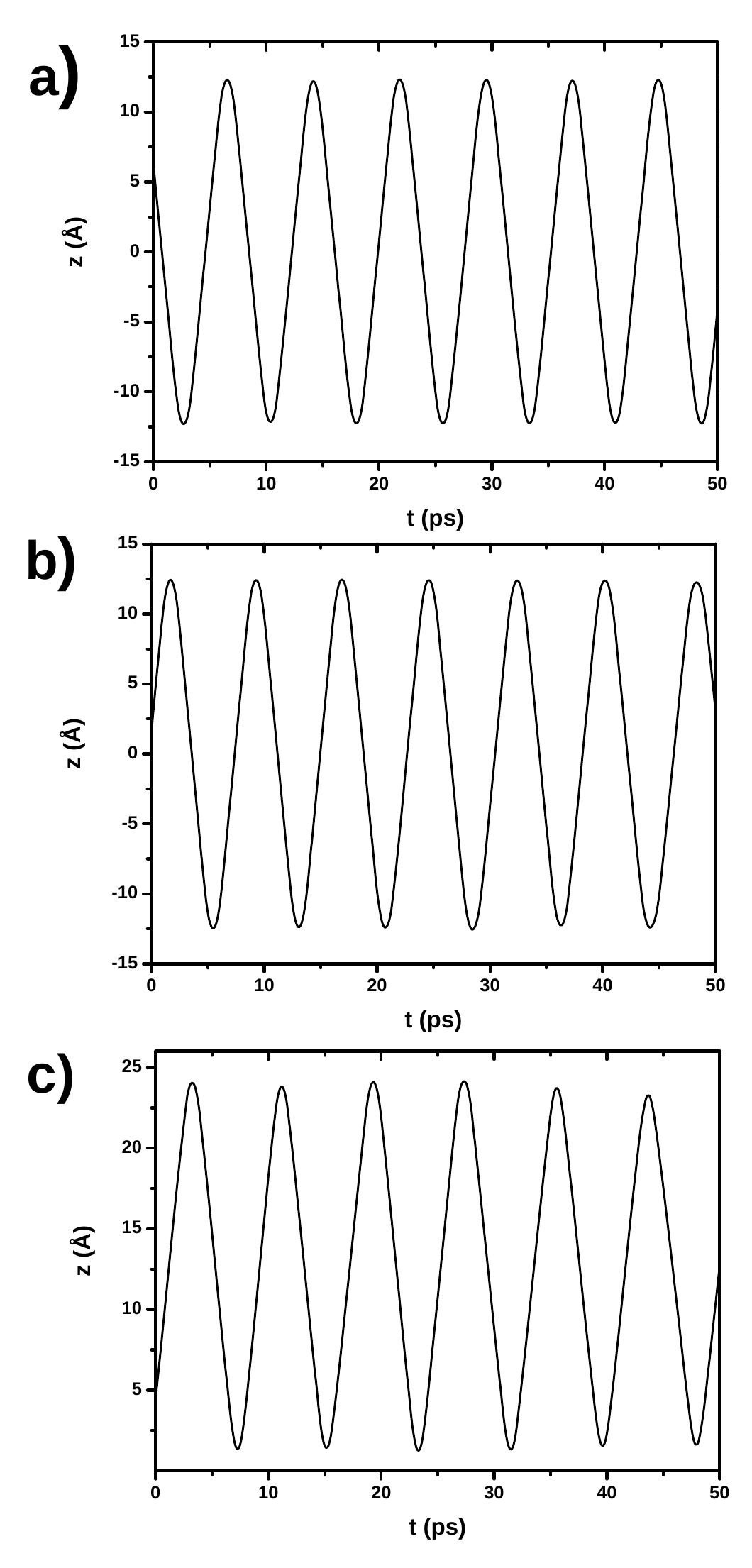}
\caption{Molecular dynamics simulation results for Ne atom
oscillations inside a \textsc{cnt} of 25{\AA} length for the
following temperatures a) 80K, b) 160K and c) 240K}
 \label{fig.lay8}
\end{center}
\end{figure}

\begin{figure}[!h]
\begin{center}
\includegraphics[scale=0.6]{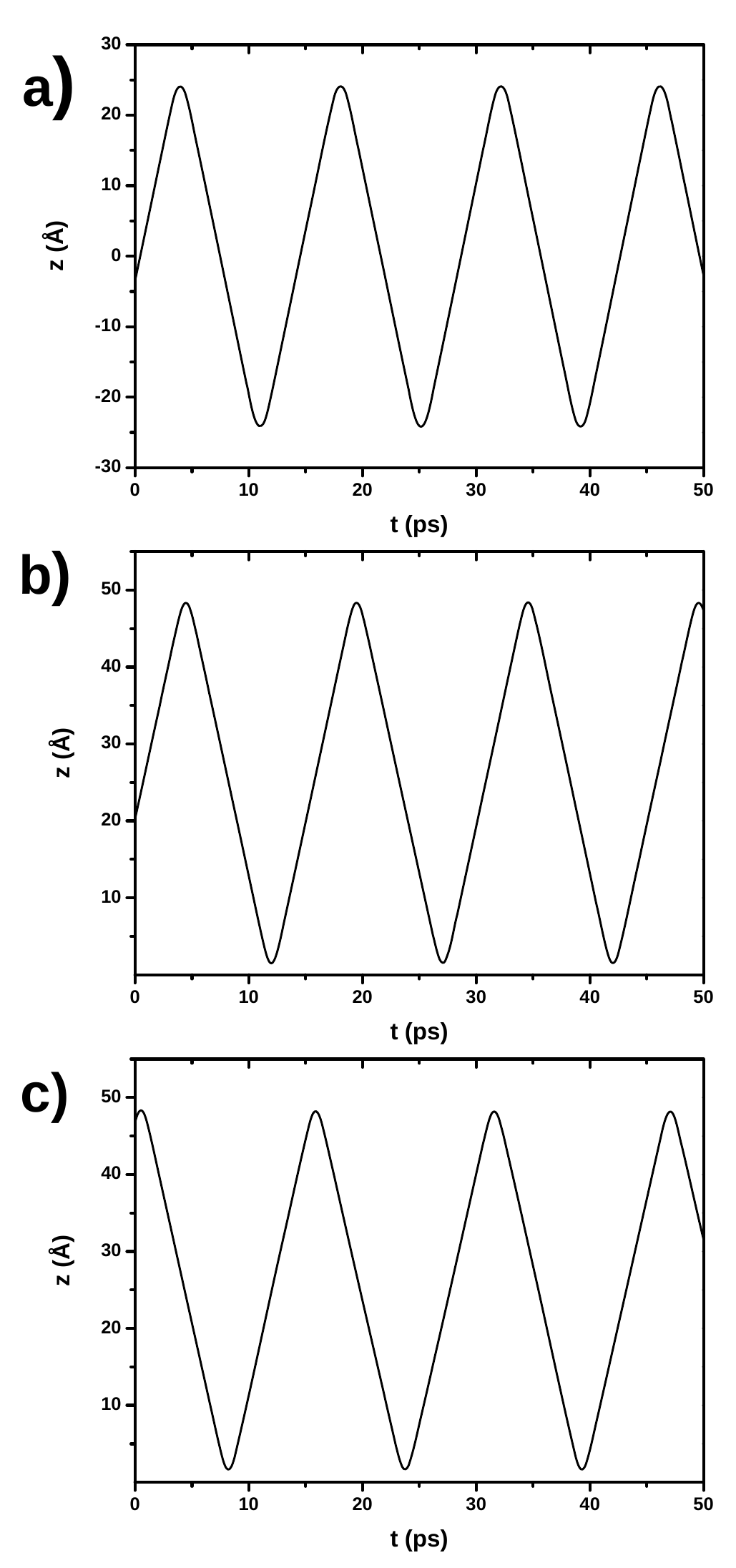}
\caption{Molecular dynamics simulation results for Ne atom
oscillations inside a \textsc{CNT} of 50{\AA} length for the
following temperatures a)80K, b)160K and c)240K}
 \label{fig.lay9}
\end{center}
\end{figure}

In Table \ref{t.Nebnnt} we present the equivalent results for the
case of a \textbf{Ne} atom inside a (5,5)\textsc{BNNT}. Again, for
the cases investigated, the temperatures where the \textbf{Ne}
atom is not encapsulated were 300K for the tube of 25 \AA, and 240
and 300 K for the tube of 50 \AA.  We believe that this occurs by
the same reasons (large end atom tube movements creating a dynamic
barrier) as in \textsc{CNT} case. That this occurs at a lower
temperature in the BNNT case can be attributed in part to
different symmetries (buckling) and the slightly diameter
differences between the CNT (6.8 \AA) and BNNT (6.9 \AA). We
expect larger fluctuations for tubes of larger lengths due to mass
differences.

\begin{table}[!h]
\caption{Calculated frequencies (\textbf{GHz}) for Ne@BNNT(5,5).}
\label{t.Nebnnt}
\begin{center}
\begin{tabular}{cccccccc}
 \hline \hline
& & & Temperature (K)& & & & \\
\hline \textbf{BNNT} &5 &10 &20 &40 &80 &160 &240 \\
\hline 25 \AA &136.71 &136.71 &136.71 &136.71 &136.71 &117.18 &117.18  \\
\hline 50 \AA &78.12  &78.12  & 78.12  &78.12  &78.12  &78.12 &- \\
\hline \hline
\end{tabular}
\end{center}
\end{table}

Again, as observed for the CNT case, sustained oscillatory regimes
are possible for a large range of temperatures, and at almost the
same frequency values. The only major differences  are at what
temperature we observe dropping frequency occurring values and the
absence of encapsulation at 240 K for the BNNT of 50 \AA. This
could be in part attribute to mass and diameter tube differences.
The major point however it is that in contrast to some previous
works \cite{lin} for the systems investigated here the symmetry
breaking and chirality of BNNT do not seem to play important roles
in determining the oscillatory regimes. The general oscillatory
behavior (Figs. 3 and 4), frequency values and axial displacement
movements (Fig. 5) are quite similar to the ones observed in the
CNT case.

\begin{figure}[!h]  
\begin{center}
\includegraphics[scale=0.6]{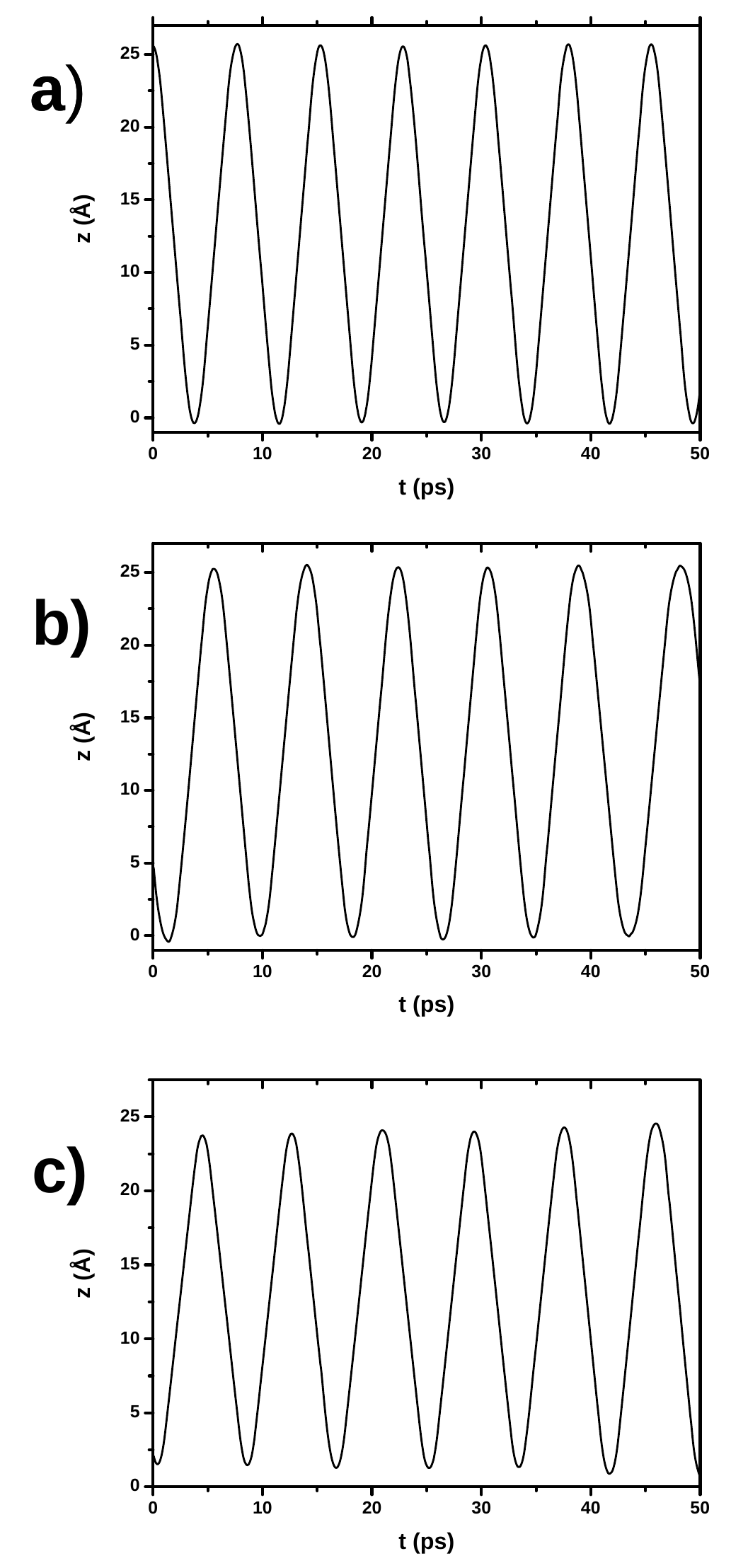}
\caption{Molecular dynamics simulation results for Ne atom
oscillations inside a \textsc{BNNT} of 25{\AA} length for the
following temperatures a)80K, b)160K and c)240K}
\label{fig.lay10}
\end{center}
\end{figure}

\begin{figure}[!h]  
\begin{center}
\includegraphics[scale=0.6]{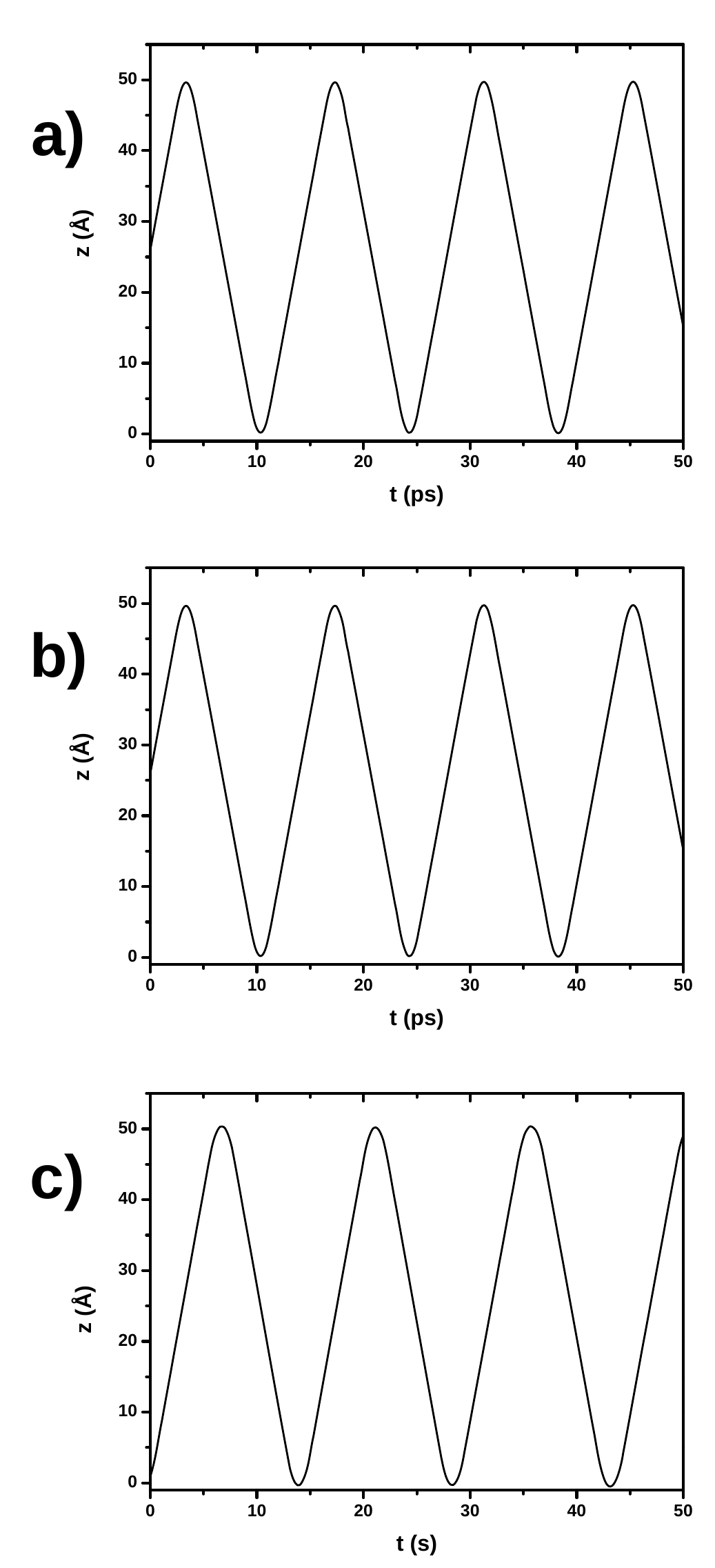}
\caption{Molecular dynamics simulation results for Ne atom
oscillations inside a \textsc{BNNT} of 50{\AA} length for the
following temperatures a)40K, b)80K and c)160K} \label{fig.lay11}
\end{center}
\end{figure}

\begin{figure}[!h]

\begin{center}
\includegraphics[scale=1.00]{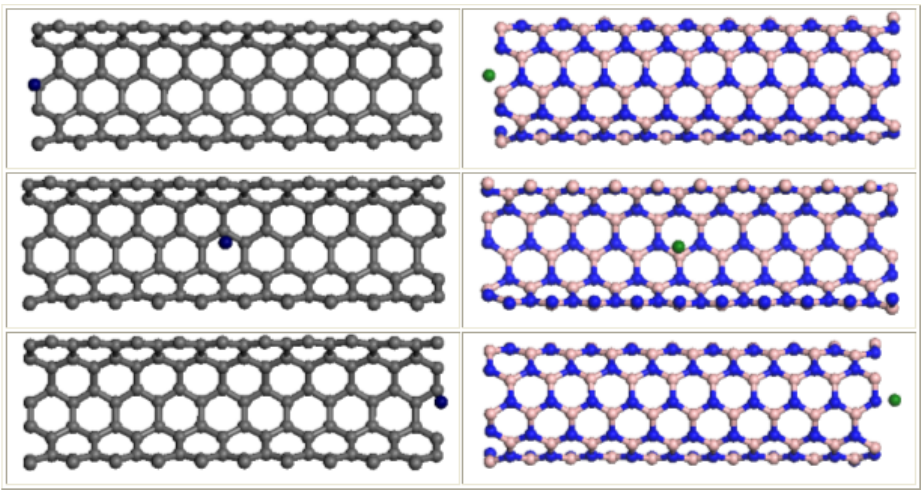}

\caption{Snapshots from the molecular dynamics simulations. Left;
Ne atom oscillating inside a \textsc{CNT} of 25{\AA} length for
the temperatures 160K. Right; Ne atom oscillating inside a
\textsc{BNNT} of same length and at the same temperature.}

 \label{figframes}
\end{center}
\end{figure}

In summary, we have carried out detailed analysis of extensive
fully atomistic molecular dynamics simulation for Ne atom
oscillating inside carbon and boron nitride nanotubes of (5,5)
chirality and different lengths (25 and 50 \AA). Our results show
that well defined sustained oscillatory regimes are possible for a
large range of temperatures (from 5 up to 240 K). There are
several stable frequencies for each one nano-oscillator
configuration studied here, ranging from 58 to 137 \textbf{GHz}.
Another interesting aspect of these nanostructures is that we can
tune the frequency by a suitable modification of the nanotube
length and/or type. Our results showed that the results for the
BNNT are quite similar to the CNT ones in terms of general
oscillatory behavior, frequency values, and axial displacement
movements. In this sense our results are closer to the obtained by
Zeng \textit{et al.} \cite{zeng} and in disagreement with the ones
obtained by Lin \textit{et al.} \cite{lin}. We hope the present
work to stimulate other studies along these lines.

\section{Acknowledgments}

This work was supported in part by IN/MCT, IMMP/MCT, THEO-NANO,
Brazilian Nanotube Network, and by the Brazilian agencies CNPq,
CAPES, FAPESP and FAPEMA.

\end{document}